\documentclass[%
reprint,
amsmath,amssymb,
aps,
prd,
showpacs
]{revtex4-1}

\usepackage{graphicx}
\usepackage{dcolumn}
\usepackage{bm}
\usepackage{dcolumn}
\usepackage{bm}
\usepackage{feynmp} 
\usepackage{nicefrac}
\usepackage{verbatim}
\usepackage{slashed}

\begin{document}
\title{Feynman Rules for QCD in Space-Cone Gauge}
\author{Alexander Karlberg}
\author{Thomas S{\o}ndergaard}
\affiliation{Niels Bohr International Academy and Niels Bohr Institute, Blegdamsvej 17, DK-2100 Copenhagen, Denmark}
\date{\today}

\begin{abstract}
  We present the Lagrangian and Feynman rules for QCD written in space-cone gauge and after eliminating unphysical degrees of freedom from the gluonic sector. The main goal is to clarify and allow for straightforward application of these Feynman rules. We comment on the connection between BCFW recursion
  relations and space-cone gauge.
\end{abstract}

\maketitle

\section{Introduction}
Calculating QCD amplitudes by means of Feynman diagrams can be an extremely challenging task. The gauge-dependence of vertices and
unphysical degrees of freedom often makes intermediate steps immensely complicated. However, in 1998 Chalmers and Siegel showed that the complexity of Feynman diagram calculations in Yang-Mills theory could be greatly reduced if a so-called space-cone gauge was used \cite{Chalmers:1998jb} (see also \cite{Chalmers:1997ui,Chalmers:2001cy} for similar simplifications in other theories).

By now several alternative approaches are also available,
such as the Britto-Cachazo-Feng-Witten (BCFW) recursion relation \cite{BCF,BCFW}.
At tree-level the above mentioned space-cone construction is closely related to these relations \cite{Vaman:2005dt}.

The main goal with this short paper is to write down all Feynman rules for QCD when working in the space-cone gauge.
To our knowledge not all of these have been explicitly presented in the literature, and it is therefore our hope that this paper will allow for easy
and straightforward application whenever such rules are needed.

The paper is structured as follows; in section \ref{sec:YML} we review the Yang-Mills Lagrangian
in space-cone gauge and the elimination of unphysical degrees of freedom. Section \ref{sec:SPF} introduces some notation and the spinor
helicity formalism. In section \ref{sec:YMF} we give the Feynman rules following from section \ref{sec:YML}. In section
\ref{sec:BCFW} we make some comments on the connection between BCFW relations and the space-cone gauge. In section \ref{sec:AF}
we add quarks to the Lagrangian and show that effective four-point vertices involving quark-antiquark pairs will appear.
In section \ref{sec:QF} we give the Feynman rules for quarks and finally in section \ref{sec:con} we have our conclusions.

\section{Yang-Mills Lagrangian in Space-cone gauge\label{sec:YML}}
We start from the standard Lagrangian of Yang-Mills theory
\begin{equation}
  \mathcal{L}_{YM} = \frac{1}{2g^2}\mathrm{Tr}[\mathcal{F}_{\mu\nu}\mathcal{F}^{\mu\nu}] = -\frac{1}{4}F^a_{\mu\nu}F^{\mu\nu\, a},
  \label{YM_lag}
\end{equation}
where
\begin{eqnarray}
  &\mathcal{F}_{\mu\nu} = -igT^aF^a_{\mu\nu}, \qquad [T^a,T^b] = if_{abc}T^c,& \nonumber \\
  &\mathrm{Tr}[T^aT^b] = \frac{1}{2}\delta^{ab},&
\end{eqnarray}
and
\begin{equation}
  F^a_{\mu\nu} = \partial_{\mu}A^a_{\nu} - \partial_{\nu}A^a_{\mu} + gf_{abc}A^b_{\mu}A^c_{\nu}.
\end{equation}
The contraction gives
\begin{align}
  F^a_{\mu\nu}F^{\mu\nu\, a} ={}& (\partial_{\mu}A^a_{\nu}-\partial_{\nu}A^a_{\mu})(\partial^{\mu}A^{\nu\,a}
  -\partial^{\nu}A^{\mu\,a}) \nonumber \\
  & +4gf_{abc}(\partial_{\mu}A^a_{\nu})A^{\mu\,b}A^{\nu\,c} \nonumber \\
  & +g^2f_{abc}f_{ab'c'}A^b_{\mu}A^c_{\nu}A^{\mu\,b'}A^{\nu\,c'}.
  \label{ff}
\end{align}
Here $A_{\mu}$ is just the usual vector field with inner product given by
\begin{equation}
  A\cdot B = A^0B^0-A^1B^1-A^2B^2-A^3B^3.
\end{equation}
We now introduce the lightcone components
\begin{align}
  A^+ \equiv \frac{1}{\sqrt{2}}(A^0+A^3), \quad A^- \equiv \frac{1}{\sqrt{2}}(A^0-A^3), \nonumber \\
  A \equiv \frac{1}{\sqrt{2}}(A^1+iA^2), \quad \bar{A} \equiv \frac{1}{\sqrt{2}}(A^1-iA^2),
\end{align}
and in terms of these the inner product is
\begin{equation}
  A\cdot B = A^+B^- + A^-B^+ - A\bar{B} - \bar{A}B.
  \label{minprod}
\end{equation}
Our first goal is to express eq.~\eqref{ff} in terms of the lightcone components, however, we use the gauge freedom to
set $A = 0$, and hence discard all terms containing an $A$. Note that in many of the intermediate calculations we have used
the fact that a term symmetric in two color indices will vanish when contracted with the anti-symmetric colorfactor.

Written in terms of the lightcone components the quadratic part of eq.~\eqref{ff} becomes
\begin{eqnarray}
  &&  (\partial_{\mu}A^a_{\nu}-\partial_{\nu}A^a_{\mu})(\partial^{\mu}A^{\nu\,a}
  -\partial^{\nu}A^{\mu\,a}) = \nonumber \\
  && 4\left[ \partial^-A^{+\,a}\partial^+A^{-\,a} - \partial A^{+\,a}\bar{\partial}A^{-\,a}
    -\bar{\partial}A^{+\,a}\partial A^{-\,a} \right. \nonumber \\
  &&\hspace{3cm} \left.+ \partial A^{+\,a}\partial^-\bar{A}^a + \partial A^{-\,a}\partial^+\bar{A}^a\right] \nonumber \\
  && -2\left[ \partial^-A^{+\,a}\partial^-A^{+\,a} + \partial^+A^{-\,a}\partial^+A^{-\,a} + \partial\bar{A}^a\partial\bar{A}^a
  \right],
\end{eqnarray}
the three-point interaction
\begin{align}
  (\partial_{\mu}A_{\nu}^a)A^{\mu\,b}A^{\nu\,c} ={}& (\partial^-A^{+\,a})A^{+\,b}A^{-\,c} + (\partial^+A^{-\,a})A^{-\,b}A^{+\,c}
  \nonumber \\
  & -(\partial A^{+\,a})\bar{A}^bA^{-\,c} - (\partial A^{-\,a})\bar{A}^bA^{+\,c},
\end{align}
and the four-point interaction
\begin{align}
  A^b_{\mu}A^c_{\nu}A^{\mu\,b'}A^{\nu\,c'} = 2A^{+\,b}A^{-\,c}A^{-\,b'}A^{+\,c'}.
\end{align}

Collecting these expressions the Lagrangian takes the form
\begin{align}
  \mathcal{L}_{YM} ={}& -\frac{1}{4}F^a_{\mu\nu}F^{\mu\nu\, a} \nonumber \\
  =& -\partial^-A^{+\,a}\partial^+A^{-\,a} + \partial A^{+\,a}\bar{\partial}A^{-\,a}
  + \bar{\partial}A^{+\,a}\partial A^{-\,a} \nonumber \\
  & - \partial A^{+\,a}\partial^-\bar{A}^a - \partial A^{-\,a}\partial^+\bar{A}^a
  \nonumber \\
  & +\frac{1}{2}\left[ \partial^-A^{+\,a}\partial^-A^{+\,a} + \partial^+A^{-\,a}\partial^+A^{-\,a} + \partial\bar{A}^a\partial\bar{A}^a
  \right] \nonumber \\
  & -g f_{abc} \left[ (\partial^-A^{+\,a})A^{+\,b}A^{-\,c} + (\partial^+A^{-\,a})A^{-\,b}A^{+\,c} \right. \nonumber \\
  &\phantom{AAAAA}\left. -(\partial A^{+\,a})\bar{A}^bA^{-\,c} - (\partial A^{-\,a})\bar{A}^bA^{+\,c}\right] \nonumber
  \\
  & -\frac{1}{2}g^2f_{abc}f_{ab'c'}A^{+\,b}A^{-\,c}A^{-\,b'}A^{+\,c'}\,.
  \label{Lbefore}
\end{align}
Following \cite{Chalmers:1998jb} we then use the equation of motion for $\bar{A}$ to eliminate this ``auxiliary'' component from the Lagrangian,
that is, we use
\begin{align}
  &\partial^+\left( \frac{\partial \mathcal{L}}{\partial (\partial^+\bar{A}^a)}\right)
  +\partial^-\left( \frac{\partial \mathcal{L}}{\partial (\partial^-\bar{A}^a)}\right) \nonumber \\
  &+\partial \left( \frac{\partial \mathcal{L}}{\partial (\partial \bar{A}^a)}\right)
  +\bar{\partial}\left( \frac{\partial \mathcal{L}}{\partial (\bar{\partial}\bar{A}^a)} \right)
  -\frac{\partial \mathcal{L}}{\partial \bar{A}^a} = 0\,,
  \label{eom}
\end{align}
and get the following expression for $\bar{A}^a$
\begin{align}
  \bar{A}^a ={}& \frac{\partial^+}{\partial}A^{-\,a} + \frac{\partial^-}{\partial}A^{+\,a} \nonumber \\
  &- gf_{abc} \frac{1}{\partial^2} \left[ ((\partial A^{+\,b})A^{-\,c}) + ((\partial A^{-\,b})A^{+\,c}) \right].
\end{align}
Plugging this back into eq.~\eqref{Lbefore}, and doing a bit of rewriting, we obtain
\begin{align}
  \mathcal{L}_{YM} ={}& A^{+\,a}\partial_{\mu}\partial^{\mu}A^{-\,a} \nonumber \\
  & +2gf_{abc}\left( \frac{\partial^-}{\partial}A^{+\,a}\right)A^{+\,b}(\partial A^{-\,c}) \nonumber \\
  &+ 2gf_{abc}\left( \frac{\partial^+}{\partial}A^{-\,b}\right)A^{-\,c}(\partial A^{+\,a}) \nonumber \\
  & +2g^2f_{abc}f_{a'bc'}\frac{1}{\partial}\left( (\partial A^{+\,a})A^{-\,c} \right)
  \frac{1}{\partial}\left( (\partial A^{-\,c'})A^{+\,a'}\right)\,.
  \label{Lafter}
\end{align}
This is the pure Yang-Mills Lagrangian written in terms of two scalar fields $A^+$ and $A^-$, consistent
with massless vector fields only having two physical degrees of freedom.

\section{Spinor formalism\label{sec:SPF}}
We choose to use the Pauli matrices with the following normalization
\begin{align}
  \sigma^0 = \frac{1}{\sqrt{2}} \left(
    \begin{array}{cc}
      1\phantom{-} & 0 \\
      0\phantom{-} & 1
    \end{array} \right), \qquad
  \sigma^1 = \frac{1}{\sqrt{2}} \left(
    \begin{array}{cc}
      0 & -1 \\
      -1 & 0
    \end{array} \right), \nonumber \\
  \sigma^2 = \frac{1}{\sqrt{2}} \left(
    \begin{array}{cc}
      0\phantom{-} & i \\
      -i\phantom{-} & 0
    \end{array} \right), \qquad
  \sigma^3 = \frac{1}{\sqrt{2}} \left(
    \begin{array}{cc}
      -1\phantom{-} & 0 \\
      0\phantom{-} & 1
    \end{array} \right),
\end{align}
such that a contraction between these and a four-vector is
\begin{align}
  P_{\dot{a}b} = P^{\mu}\sigma_{\mu} = \frac{1}{\sqrt{2}}\left(
    \begin{array}{cc}
      p^0+p^3 &\phantom{A} p^1-ip^2 \\
      p^1+ip^2 &\phantom{A} p^0-p^3
    \end{array} \right)
  =
  \left(
    \begin{array}{cc}
      p^+ &\phantom{a} \bar{p}\phantom{^-} \\
      p\phantom{^+} &\phantom{a} p^-
    \end{array} \right),
  \label{lform}
\end{align}
and
\begin{align}
  P^{\dot{a}b} =\left(
    \begin{array}{cc}
      p^- & -p\phantom{^-} \\
      -\bar{p}\phantom{^+} & p^+
    \end{array} \right).
\end{align}
Note that $\det(P) = \frac{1}{2}P^2$, so if $P^2 = 0$ this matrix only has one non-vanishing eigenvalue and can be decomposed into
a bispinor product
\begin{align}
  P_{\dot{a}b} = p_{\dot{a}}p_{b}\,.
\end{align}
We will from now on use the following braket notation
\begin{align}
  p_a \equiv |p\rangle, \quad p_{\dot{a}} \equiv \lbrack p|, \quad p^a \equiv \langle p|, \quad p^{\dot{a}} \equiv |p\rbrack\,.
\end{align}
Notice that our convention differs from \cite{Chalmers:1998jb}.

\subsection{Reference frame}
As always when one wants to do amplitude calculations by Feynman rules there are some Lorentz frames in which the
calculations are easier to perform. To set up this we introduce two (for now) arbitrary massless \textit{reference} momenta, one denoted with a $\oplus$ and
one denoted with $\ominus$.
We then choose to work in the Lorentz frame where our two reference momenta have the following simple expressions
\begin{align}
  P_{\oplus} = \left(
    \begin{array}{cc}
      0 &\phantom{a} 0 \\
      0 &\phantom{a} p^-_{\oplus}
    \end{array}\right)
  = |-\rangle\lbrack -|
  = \left( \begin{array}{c}
      0 \\
      \sqrt{p^-_{\oplus}}
    \end{array} \right)
  \left( \begin{array}{cc}
      0 & \sqrt{p^-_{\oplus}}
    \end{array} \right),
  \\
  P_{\ominus} = \left(
    \begin{array}{cc}
      p^+_{\ominus} &\phantom{a} 0 \\
      0 &\phantom{a} 0
    \end{array}\right)
  = |+\rangle\lbrack +|
  = \left( \begin{array}{c}
      \sqrt{p^+_{\ominus}} \\
      0
    \end{array} \right)
  \left( \begin{array}{cc}
      \sqrt{p^+_{\ominus}} & 0
    \end{array} \right),
\end{align}
that is the frame where they both move along the $i=3$ axis, but in opposite direction. Note that
$P_{\oplus}\cdot P_{\ominus} = p^-_{\oplus}p^+_{\ominus} = \langle + - \rangle \lbrack -+\rbrack$, and that
we have called the spinor of the $\oplus$ momentum for $|-\rangle$ and vice versa. Our choice of labelling
will soon become apparent. Since any four-vector,
contracted with the Pauli matrices, is written in the form of eq.~\eqref{lform}, using the following normalized matrices
as basis
\begin{align}
  \frac{|+\rangle\lbrack +|}{p^+_{\ominus}}, \quad \frac{|-\rangle\lbrack -|}{p^-_{\oplus}}, \quad
  \frac{|+\rangle\lbrack -|}{\sqrt{\langle +-\rangle\lbrack -+\rbrack}}, \quad
  \frac{|-\rangle\lbrack +|}{\sqrt{\langle +-\rangle\lbrack -+\rbrack}},
\end{align}
four-vectors take the form
\begin{align}
  P ={}& p^+\frac{|+\rangle\lbrack +|}{p^+_{\ominus}} + p^-\frac{|-\rangle\lbrack -|}{p^-_{\oplus}} \nonumber \\
  & +\bar{p}\frac{|+\rangle\lbrack -|}{\sqrt{\langle +-\rangle\lbrack -+\rbrack}}
  +p\frac{|-\rangle\lbrack +|}{\sqrt{\langle +-\rangle\lbrack -+\rbrack}}\,,
  \label{newbasis}
\end{align}
where the coefficients are just the lightcone components. Since $P = |p\rangle\lbrack p|$ the lightcone components are
\begin{align}
  p^+ = \frac{\langle -p\rangle\lbrack p-\rbrack}{p^-_{\oplus}}\,, \quad
  p^- = \frac{\langle +p\rangle\lbrack p+\rbrack}{p^+_{\ominus}}\,, \nonumber \\
  \bar{p} = \frac{-\langle -p\rangle\lbrack p+\rbrack}{\sqrt{\langle +-\rangle\lbrack -+\rbrack}}\,, \quad
  p = \frac{-\langle +p\rangle\lbrack p-\rbrack}{\sqrt{\langle +-\rangle\lbrack -+\rbrack}}\,.
  \label{lc_exp}
\end{align}

We also choose to use our reference momenta in the expression for the polarization vectors $\epsilon_{\pm}(P)$ 
\begin{align}
  \epsilon_+(P) = \frac{|+\rangle\lbrack p|}{\langle +p\rangle}\,, \qquad
  \epsilon_-(P) = \frac{|p\rangle\lbrack -|}{\lbrack p-\rbrack}\,.
\end{align}
That is, we have used our reference momenta to fix the gauge-freedom one has in polarization vectors.

From eq.~\eqref{Lafter} it is evident that we are only concerned with the $(\epsilon)^{\pm}$ components, and from eq.~\eqref{newbasis}
we see that
\begin{align}
  (\epsilon_+)^+ &= \frac{\langle -+\rangle\lbrack p-\rbrack}{\langle +p\rangle p^-_{\oplus}}\,, \qquad (\epsilon_+)^- = 0\,, \\
  (\epsilon_-)^- &= \frac{\langle +p\rangle\lbrack -+\rbrack}{\lbrack p-\rbrack p^+_{\ominus}}\,, \qquad (\epsilon_-)^+ = 0\,.
\end{align}
Hence, with this setup we have that only the positive helicity gluons have the $+$ lightcone component and only the negative helicity gluons
the $-$ component. For this reason the $\pm$ labels in eq.~\eqref{Lafter} is now actually denoting the helicity and
not just the specific lightcone component.

\section{Feynman rules in pure Yang-Mills\label{sec:YMF}}
In this section we present the color-ordered Feynman rules one obtains from eq.~\eqref{Lafter}, that is the rules one could use in calculating,
for instance, the partial tree amplitudes $A_n$ in
\begin{align}
\mathcal{A}_n = 2 g^{n-2} \sum \mathrm{Tr}[T^{a_1}T^{a_2}\cdots T^{a_n}] A_n(1,2,\ldots,n)\,,
\end{align}
where the sum is over all non-cyclic permutations of external legs.
Before we do so, remember that in the last section our massless reference
momenta was just arbitrarily chosen, however, if we make the choice that $P_{\oplus}$ is one of the \textit{external} momenta of a $+$ helicity gluon
and $P_{\ominus}$ one of the \textit{external} momenta of a $-$ helicity gluon, the rules and explicit calculations simplify greatly. 

The external \textit{non}-reference legs will just contribute with the plus or
minus lightcone component of the polarization vector, depending on the helicity (note that we \textit{always} take
external momenta to be outgoing), \textit{i.e.}
\begin{eqnarray}
  \parbox{30mm}{
    \begin{fmffile}{ex.plus}
      \begin{fmfgraph*}(60,40)
        \fmfleft{i1}
        \fmfright{o1}
        \fmf{gluon}{i1,o1}

        \fmfdot{o1}

        \fmflabel{$P^+$}{i1}
      \end{fmfgraph*}
    \end{fmffile}
  }
  \!\!\!\!\! =
  (\epsilon_+(P))^+ \,,
\end{eqnarray}
or
\begin{eqnarray}
  \parbox{30mm}{
    \begin{fmffile}{ex.minus}
      \begin{fmfgraph*}(60,40)
        \fmfleft{i1}
        \fmfright{o1}
        \fmf{gluon}{i1,o1}

        \fmfdot{o1}

        \fmflabel{$P^-$}{i1}
      \end{fmfgraph*}
    \end{fmffile}
  }
  \!\!\!\!\! =
  (\epsilon_-(P))^- \,.
\end{eqnarray}
However, for the reference legs the $(\epsilon_{\pm})^{\pm}$ vanish because of the
$\lbrack p-\rbrack$ and $\langle +p\rangle$ in the numerator. These factors can only be countered in the three-point vertex and
only if they sit on the $\partial^{\pm}A^{\mp}/ \partial$ term, \textit{i.e.}
\begin{align}
  \frac{p^-}{p}(\epsilon_+)^+ =
  \frac{\lbrack p+\rbrack \sqrt{\langle +-\rangle \lbrack -+\rbrack}}{\langle +p\rangle\lbrack -+\rbrack}\quad
  \xrightarrow{p\rightarrow -}  \quad \frac{\lbrack -+\rbrack}{\sqrt{\langle +-\rangle \lbrack -+\rbrack}}, \nonumber \\
  \frac{p^+}{p}(\epsilon_-)^- =
  \frac{-\langle -p\rangle \lbrack -+\rbrack}{\lbrack p-\rbrack \sqrt{\langle +-\rangle \lbrack -+\rbrack}} \quad
  \xrightarrow{p\rightarrow +} \quad \frac{\langle -+\rangle}{\sqrt{\langle +-\rangle\lbrack -+\rbrack}}.
  \label{cancellation}
\end{align}
The product of these two contributions is $-1$, and since every diagram will always contain this product
(assuming one takes both reference momenta to correspond to external legs),
we just write the external lines for reference legs as
\begin{equation}
  \parbox{30mm}{
    \begin{fmffile}{ex.ref}
      \begin{fmfgraph*}(60,40)
        \fmfleft{i1}
        \fmfright{o1}
        \fmf{gluon}{i1,o1}

        \fmfdot{o1}

        \fmflabel{$P^{\pm}_{ref}$}{i1}
      \end{fmfgraph*}
    \end{fmffile}
  }
  \!\!\!\!\! =
  i\,.
  \label{ex.ref}
\end{equation}

The term representing the propagator is the usual boson propagator (notice that we for simplicity discard all factors of $i$ in the following rules)
\begin{equation}
  \parbox{30mm}{
    \begin{fmffile}{prop}
      \begin{fmfgraph*}(60,40)
        \fmfleft{i1}
        \fmfright{o1}
        \fmf{gluon,label=$Q$,l.side=left}{i1,o1}

        \fmfdot{i1,o1}

      \end{fmfgraph*}
    \end{fmffile}
  }
  \!\!\!\!\! =
  \frac{1}{Q^2}\,.
\end{equation}

The three-point vertex splits into the case where one of the lines is a reference leg and the case in which non of the lines
are. With a reference leg present we have used up the $\partial^{\pm}A^{\mp}/ \partial$ term for a cancellation like in eq.~\eqref{cancellation}
and are only left with the contribution from the opposite helicity leg, through $\partial A^{\pm}$, \textit{i.e.} \\
\begin{eqnarray}
  \parbox{30mm}{
    \begin{fmffile}{3vref1}
      \begin{fmfgraph*}(70,50)
        \fmfleft{i1,i2}
        \fmfright{o1}
        \fmf{gluon}{i1,v1}
        \fmf{gluon}{i2,v1}
        \fmf{gluon}{o1,v1}

        \fmflabel{$P^{\pm}_{ref}$}{i1}
        \fmflabel{$K^{\pm}$}{i2}
        \fmflabel{$Q^{\mp}$}{o1}
      \end{fmfgraph*}
    \end{fmffile}
  }
  \phantom{aa} = \phantom{a} 2q \,,
  \label{3vref1}
\end{eqnarray}
or
\\
\begin{eqnarray}
  \parbox{30mm}{
    \begin{fmffile}{3vref2}
      \begin{fmfgraph*}(70,50)
        \fmfleft{i1,i2}
        \fmfright{o1}
        \fmf{gluon}{i1,v1}
        \fmf{gluon}{i2,v1}
        \fmf{gluon}{o1,v1}
        \fmflabel{$P^{\pm}$}{i1}
        \fmflabel{$K^{\pm}_{ref}$}{i2}
        \fmflabel{$Q^{\mp}$}{o1}
      \end{fmfgraph*}
    \end{fmffile}}
  \phantom{aa} = \phantom{a} -2q\,,
  \label{3vref2}
\end{eqnarray}
\\ \\
where the minus sign in the second diagram comes from the antisymmetry of the colorfactor $f_{abc}$. 
There can only be \textit{one} reference leg on a three-point vertex since there is only one term in these
that can counter the vanishing of that leg.

In the second case of no reference leg, the three-point vertex is
\\
\begin{equation}
  \parbox{30mm}{
    \begin{fmffile}{3v}
      \begin{fmfgraph*}(70,50)
        \fmfleft{i1,i2}
        \fmfright{o1}
        \fmf{gluon}{i1,v1}
        \fmf{gluon}{i2,v1}
        \fmf{gluon}{o1,v1}

        \fmflabel{$P^{\pm}$}{i1}
        \fmflabel{$K^{\pm}$}{i2}
        \fmflabel{$Q^{\mp}$}{o1}
      \end{fmfgraph*}
    \end{fmffile}
  }
  \phantom{aa} = \phantom{a}2
  q\left(\frac{p^{\mp}}{p} - \frac{k^{\mp}}{k} \right) \,.
\end{equation}
\\ \\
Again the minus sign is a consequence of the anti-symmetric colorfactor.

We can never have a reference leg on a four-point vertex since these do not contribute with a similar ``$1/0$'' counter-term. However, there are still two different kinds of four-point vertices because of the two possible helicity
configurations $++--$ and $-+-+$ (cyclically speaking). The first case is \\
\begin{equation}
  \parbox{30mm}{
    \begin{fmffile}{4v1}
      \begin{fmfgraph*}(70,50)
        \fmfleft{i1,i2}
        \fmfright{o1,o2}
        \fmf{gluon}{i1,v1}
        \fmf{gluon}{i2,v1}
        \fmf{gluon}{o1,v1}
        \fmf{gluon}{o2,v1}

        \fmflabel{$P^{+}$}{i1}
        \fmflabel{$K^{-}$}{i2}
        \fmflabel{$Q^{-}$}{o2}
        \fmflabel{$T^{+}$}{o1}
      \end{fmfgraph*}
    \end{fmffile}
  }
  = \phantom{a} -2\frac{pq+tk}{(p+k)^2} \,,
\end{equation}
\\ \\
and the second case
\\ \\
\begin{equation}
  \parbox{30mm}{
      \begin{fmffile}{4v2}
        \begin{fmfgraph*}(70,50)
          \fmfleft{i1,i2}
          \fmfright{o1,o2}
          \fmf{gluon}{i1,v1}
          \fmf{gluon}{i2,v1}
          \fmf{gluon}{o1,v1}
          \fmf{gluon}{o2,v1}
          \fmflabel{$P^{+}$}{i1}
          \fmflabel{$K^{-}$}{i2}
          \fmflabel{$Q^{-}$}{o1}
          \fmflabel{$T^{+}$}{o2}
        \end{fmfgraph*}
      \end{fmffile}
  }
  = \phantom{a} 2\frac{pk+tq}{(p+q)^2} + 2\frac{pq+kt}{(p+k)^2} \,.
\end{equation} \\

These rules reduce the number of diagrams contributing to a specific color-ordered amplitude significantly. However,
the reduction relies heavily on the choice of having external momenta as reference momenta such that diagrams with
both reference legs on the same three-point vertex and diagrams with a reference leg on a four-point vertex all vanish. The rules are
of course perfectly allowed without this choice, but then no such constraints exist and the simplified diagrams \eqref{ex.ref}, \eqref{3vref1} and
\eqref{3vref2} should be discarded.

Before turning to the inclusion of quarks let us make some comments on the connection between working in space-cone gauge and
the BCFW recursion relation.


\section{BCFW relations from space-cone gauge\label{sec:BCFW}}
One of the interesting features of the space-cone gauge is, that a diagrammatic proof of the BCFW recursion relation can be obtained from the above Feynman rules. The main observation is that no vertex depends on $\bar{p}$. Therefore, a shift in the $|+\rangle\lbrack -|$ direction of a reference leg leaves no imprint on the vertices, but only on internal propagators. Using a propagator relation between shifted and unshifted propagators, it is possible to relate amplitudes calculated in space-cone gauge to the BCFW result. For the four- and five-point case this calculation was done in \cite{Vaman:2005dt} where the reader may also find details regarding the propagator identity. However, these two examples are in a way special, since no four-point vertex enters the calculations. The first case where the four-point vertex is manifest for all choices of reference momenta is the $NMHV$ six-point amplitude. In a very condensed notation it is given by
\begin{align} &A_6(\oplus ++-- \ominus) =
  \parbox{30mm}{\fmfframe(0,0)(-30,0){\begin{fmffile}{6p1}
        \begin{fmfgraph*}(40,30)
          \fmfstraight
          \fmftop{t1,t2,t3,t4}
          \fmfbottom{b1,b2,b3,b4}
          \fmf{plain,tension=0.5}{t1,v1}
          \fmf{plain,tension=0.5}{b1,v1}
          \fmf{plain}{v1,v2}
          \fmf{plain}{t2,v2}
          \fmf{plain}{v2,v3}
          \fmf{plain}{t3,v3}
          \fmf{plain}{v3,v4}
          \fmf{plain,tension=0.5}{t4,v4}
          \fmf{plain,tension=0.5}{b4,v4}
          \fmf{phantom}{b2,v2}
          \fmf{phantom}{b3,v3}
        \end{fmfgraph*}
      \end{fmffile}}} \hspace{-11mm}+
  \parbox{30mm}{\fmfframe(0,0)(0,0){\begin{fmffile}{6p2}
        \begin{fmfgraph*}(40,30)
          \fmfstraight
          \fmftop{t1,t2,t3,t4}
          \fmfbottom{b1,b2,b3,b4}
          \fmf{plain,tension=0.5}{t1,v1}
          \fmf{plain,tension=0.5}{b1,v1}
          \fmf{plain}{v1,v2}
          \fmf{plain}{t2,v2}
          \fmf{plain}{v2,v3}
          \fmf{phantom}{t3,v3}
          \fmf{plain}{v3,v4}
          \fmf{plain,tension=0.5}{t4,v4}
          \fmf{plain,tension=0.5}{b4,v4}
          \fmf{phantom}{b2,v2}
          \fmf{plain}{b3,v3}
        \end{fmfgraph*}
      \end{fmffile}}} \nonumber \\ &+
  \parbox{30mm}{\fmfframe(0,0)(0,0){\begin{fmffile}{6p3}
        \begin{fmfgraph*}(30,30)
          \fmfsurroundn{e}{6}
          \fmf{plain}{e1,v1}
          \fmf{plain}{e2,v1}
          \fmf{plain}{v1,v4}
          \fmf{plain}{e3,v2}
          \fmf{plain}{e4,v2}
          \fmf{plain}{v2,v4}
          \fmf{plain}{e5,v3}
          \fmf{plain}{e6,v3}
          \fmf{plain}{v3,v4}
        \end{fmfgraph*}
      \end{fmffile}}}\hspace{-12mm}+
  \parbox{30mm}{\fmfframe(0,0)(-30,0){\begin{fmffile}{6p4}
        \begin{fmfgraph*}(40,30)
          \fmfstraight
          \fmftop{t1,t2,t3,t4}
          \fmfbottom{b1,b2,b3,b4}
          \fmf{plain,tension=0.5}{t1,v1}
          \fmf{plain,tension=0.5}{b1,v1}
          \fmf{plain}{v1,v2}
          \fmf{plain}{t2,v2}
          \fmf{plain}{t3,v2}
          \fmf{plain}{v2,v3}
          \fmf{plain,tension=0.5}{t4,v3}
          \fmf{plain,tension=0.5}{b4,v3}
          \fmf{phantom}{b2,v2}
          \fmf{phantom}{b3,v2}
        \end{fmfgraph*}
      \end{fmffile}}} \hspace{-11mm}+
  \parbox{30mm}{\fmfframe(0,0)(-30,0){\begin{fmffile}{6p5}
        \begin{fmfgraph*}(40,30)
          \fmfstraight
          \fmftop{t1,t2,t3}
          \fmfbottom{b1,b2,b3}
          \fmfright{e1}
          \fmfleft{p1}
          \fmf{plain,tension=0.5}{t1,v1}
          \fmf{plain,tension=0.5}{b1,v1}
          \fmf{plain}{v1,v2}
          \fmf{plain}{t2,v2}
          \fmf{plain}{v2,v3}
          \fmf{plain,tension=0.5}{t3,v3}
          \fmf{plain,tension=0.5}{b3,v3}
          \fmf{phantom}{b2,v2}
          \fmf{plain,tension=0}{e1,v3}
          \fmf{phantom,tension=0}{p1,v1}
        \end{fmfgraph*}
      \end{fmffile}}} \hspace{-10mm}
  \label{sixpoint}
\end{align}
where $\oplus$ and $\ominus$ denotes the positive and negative reference gluons respectively. When we use the propagator identity introduced by Vaman and Yao, we obtain a factorization which has the structure of a three-point amplitude times a five-point amplitude. Here the four-point vertex enters explicitly in the five-point amplitude, which seems to be in conflict with \cite{Chalmers:1998jb}, where it was shown explicitly that the five-point amplitude is independent of the four-point vertex.
\vspace{5mm}
\begin{equation}
  \parbox{30mm}{\fmfframe(0,0)(-30,0){\begin{fmffile}{vaman1}
        \begin{fmfgraph*}(40,30)
          \fmfstraight
          \fmftop{t1,t2,t3}
          \fmfbottom{b1,b2,b3}
          \fmfright{e1}
          \fmfleft{p1}
          \fmf{plain,tension=0.5}{t1,v1}
          \fmf{plain,tension=0.5}{b1,v1}
          \fmf{plain}{v1,v2}
          \fmf{plain}{t2,v2}
          \fmf{plain}{v2,v3}
          \fmf{plain,tension=0.5}{t3,v3}
          \fmf{plain,tension=0.5}{b3,v3}
          \fmf{phantom}{b2,v2}
          \fmf{plain,tension=0}{e1,v3}
          \fmf{phantom,tension=0}{p1,v1}
          \fmflabel{$\oplus$}{t2}
          \fmflabel{$+$}{t3}
          \fmflabel{$+$}{e1}
          \fmflabel{$-$}{b3}
          \fmflabel{$-$}{b1}
          \fmflabel{$\ominus$}{t1}
        \end{fmfgraph*}
      \end{fmffile}}} \hspace{-5mm} \rightarrow \hspace{5mm}
  \parbox{30mm}{\fmfframe(0,0)(-30,0){\begin{fmffile}{vaman1_3}
        \begin{fmfgraph*}(17,30)
          \fmfstraight
          \fmfleft{l1,l2}
          \fmfright{e1}
          \fmf{plain,tension=0.5}{l1,v1}
          \fmf{plain,tension=0.5}{l2,v1}
          \fmf{plain}{v1,e1}
          \fmflabel{$-$}{l1}
          \fmflabel{$\hat{\ominus}$}{l2}
        \end{fmfgraph*}
      \end{fmffile}}} \hspace{-20mm}\frac{1}{P^2}
  \parbox{30mm}{\fmfframe(0,0)(0,0){\begin{fmffile}{vaman1_5}
        \begin{fmfgraph*}(30,30)
          \fmfstraight
          \fmfleft{l1}
          \fmfright{e1}
          \fmftop{t1,t2,t3}
          \fmfbottom{b1,b2,b3}
          \fmf{plain,tension=0}{l1,v1}
          \fmf{phantom,tension=0.5}{t1,v1}
          \fmf{phantom,tension=0.5}{b1,v1}
          \fmf{plain}{t2,v1}
          \fmf{phantom}{b2,v1}
          \fmf{plain,tension=1.5}{v1,v2}
          \fmf{plain}{v2,e1}
          \fmf{plain,tension=0.5}{t3,v2}
          \fmf{plain,tension=0.5}{b3,v2}
          \fmflabel{$\hat{\oplus}$}{t2}
          \fmflabel{$+$}{t3}
          \fmflabel{$+$}{e1}
          \fmflabel{$-$}{b3}
        \end{fmfgraph*}
      \end{fmffile}}}
  \label{sixpointvaman1}
\end{equation}
\begin{equation}
  \parbox{30mm}{\fmfframe(0,20)(0,20){\begin{fmffile}{vaman2}
        \begin{fmfgraph*}(40,30)
          \fmfstraight
          \fmftop{t1,t2,t3,t4}
          \fmfbottom{b1,b2,b3,b4}
          \fmf{plain,tension=0.5}{t1,v1}
          \fmf{plain,tension=0.5}{b1,v1}
          \fmf{plain}{v1,v2}
          \fmf{plain}{t2,v2}
          \fmf{plain}{t3,v2}
          \fmf{plain}{v2,v3}
          \fmf{plain,tension=0.5}{t4,v3}
          \fmf{plain,tension=0.5}{b4,v3}
          \fmf{phantom}{b2,v2}
          \fmf{phantom}{b3,v2}
          \fmflabel{$+$}{t1}
          \fmflabel{$+$}{t2}
          \fmflabel{$-$}{t3}
          \fmflabel{$-$}{t4}
          \fmflabel{$\oplus$}{b1}
          \fmflabel{$\ominus$}{b4}
        \end{fmfgraph*}
      \end{fmffile}}}\hspace{-5mm} \rightarrow \hspace{5mm}
  \parbox{30mm}{\fmfframe(0,20)(0,20){\begin{fmffile}{vaman2_5}
        \begin{fmfgraph*}(30,30)
          \fmfstraight
          \fmfleft{l1}
          \fmfright{e1}
          \fmftop{t1,t2,t3}
          \fmfbottom{b1,b2,b3}
          \fmf{plain}{t1,v1}
          \fmf{plain}{b1,v1}
          \fmf{phantom}{l1,v1}
          \fmf{plain}{v1,v2}
          \fmf{plain}{t2,v2}
          \fmf{phantom}{b2,v2}
          \fmf{plain}{e1,v2}
          \fmf{plain}{t3,v2}
          \fmf{phantom}{b3,v2}
          \fmflabel{$\hat{\oplus}$}{b1}
          \fmflabel{$+$}{t1}
          \fmflabel{$+$}{t2}
          \fmflabel{$-$}{t3}
        \end{fmfgraph*}
      \end{fmffile}}}\hspace{-15mm}\frac{1}{P^2}
  \parbox{30mm}{\fmfframe(0,20)(0,20){\begin{fmffile}{vaman2_3}
        \begin{fmfgraph*}(17,30)
          \fmfstraight
          \fmfleft{l1}
          \fmfright{e1,e2}
          \fmf{plain,tension=0.5}{v1,e1}
          \fmf{plain,tension=0.5}{v1,e2}
          \fmf{plain}{l1,v1}
          \fmflabel{$-$}{e2}
          \fmflabel{$\hat{\ominus}$}{e1}
        \end{fmfgraph*}
      \end{fmffile}}} \hspace{-10mm}
  \label{sixpointvaman2}
\end{equation}
To explain this, we notice, that the result in \cite{Chalmers:1998jb} exploits the fact that in the space-cone gauge, the number of diagrams and the types of vertices that enters a specific calculation, is highly dependent on the chosen reference lines. When two of the external gluons are chosen as references, the four-point vertex in the five-point amplitude is absent. However, if we only choose one of the external momenta as reference, the four-point vertex will still be present in the calculation. The structure of the five-point amplitude, with only one external gluon chosen as reference line, is shown below.
\begin{align}
  A_5(\oplus++--)=&
  \parbox{30mm}{\fmfframe(0,5)(0,0){\begin{fmffile}{5p1}
        \begin{fmfgraph*}(30,30)
          \fmfstraight
          \fmftop{t1,t2,t3}
          \fmfbottom{b1,b2,b3}
          \fmf{plain}{t1,v1}
          \fmf{plain}{b1,v1}
          \fmf{plain}{v1,v2}
          \fmf{plain}{t2,v2}
          \fmf{phantom}{b2,v2}
          \fmf{plain}{v2,v3}
          \fmf{plain}{t3,v3}
          \fmf{plain}{b3,v3}
        \end{fmfgraph*}
      \end{fmffile}}} \hspace{-10mm} +\hspace{5mm}
  \parbox{30mm}{\fmfframe(0,5)(0,0){\begin{fmffile}{5p2}
        \begin{fmfgraph*}(30,30)
          \fmfstraight
          \fmftop{t1,t2}
          \fmfbottom{b1,b2}
          \fmfleft{l1}
          \fmfright{r1}
          \fmf{plain}{t1,v1}
          \fmf{plain}{b1,v1}
          \fmf{plain,tension=1.3}{v1,v2}
          \fmf{plain}{t2,v2}
          \fmf{plain}{b2,v2}
          \fmf{plain}{v2,r1}
          \fmf{phantom}{l1,v1}
          \fmflabel{$\oplus$}{t1}
          \fmflabel{$+$}{t2}
          \fmflabel{$+$}{r1}
          \fmflabel{$-$}{b1}
          \fmflabel{$-$}{b2}
        \end{fmfgraph*}
      \end{fmffile}}} \nonumber \\ & + \hspace{10mm}
  \parbox{30mm}{\fmfframe(0,20)(0,20){\begin{fmffile}{5p3}
        \begin{fmfgraph*}(30,30)
          \fmfstraight
          \fmftop{t1,t2}
          \fmfbottom{b1,b2}
          \fmfleft{l1}
          \fmfright{r1}
          \fmf{plain}{t1,v1}
          \fmf{plain}{b1,v1}
          \fmf{plain,tension=1.3}{v1,v2}
          \fmf{plain}{t2,v2}
          \fmf{plain}{b2,v2}
          \fmf{plain}{v2,r1}
          \fmf{phantom}{l1,v1}
          \fmflabel{$+$}{t1}
          \fmflabel{$+$}{t2}
          \fmflabel{$-$}{r1}
          \fmflabel{$-$}{b2}
          \fmflabel{$\oplus$}{b1}
        \end{fmfgraph*}
      \end{fmffile}}} \hspace{-5mm}
  \label{fivepoints}
\end{align}
Comparing this with \eqref{sixpointvaman1} and \eqref{sixpointvaman2} we notice, that the diagrams with four-point vertices exactly matches those of \eqref{fivepoints}. The diagrams with only three-point vertices can just as easily be found when using the propagator identity in the rest of the diagrams. We therefore conclude that the amplitudes combine in exactly the right way as to reproduce the BCFW result, and that we should expect the procedure to generalize to any number of external legs.

Lastly, we would like to make a comment regarding the propagator identity that is needed in order to conclude the BCFW recursion relation in \cite{Vaman:2005dt}. It is shown that the identity is equivalent to
\begin{equation}
  \int \frac{\mathrm{d}z}{z(z-z_1)(z-z_2)\cdots(z-z_{n-1})}=0\,,
  \label{contour}
\end{equation}
over a contour enclosing all poles. In the original proof of the BCFW recursion relation \cite{BCFW}, two main properties of tree-level amplitudes was needed. First that the poles of the shifted amplitude $A(z)$ all come from propagators going on-shell. Second, that $A(z)\rightarrow 0$ as $z\rightarrow \infty$. These statements are proved in \cite{BCFW}, but the arguments become almost trivial when put into the light of the space-cone formalism.

The first statement is immediately clear since the shifts can be chosen such that the vertices of the amplitude are unaffected. Hence, only internal propagators change. Then the second statement follows from \eqref{contour} since the integrand obviously converges fast enough to zero.

\section{Adding fermions\label{sec:AF}}
We will now add quarks to our Lagrangian. This amounts to adding
\begin{align}
  \mathcal{L}_{q} &= \bar{\psi}(i\gamma_{\mu}D^{\mu})\psi - m \bar{\psi}\psi \nonumber \\
  &=  \mathcal{L}_{q,0} + gA^{a\mu} \bar{\psi} \gamma_{\mu} T^a\psi \,,
\end{align}
to eq.~\eqref{YM_lag}, where $\mathcal{L}_{q,0}$ is just the Lagrangian for the free Dirac field, and the
interaction can be written out in terms of the lightcone components as
\begin{align}
  gA^{a\mu} \bar{\psi} \gamma_{\mu} T^a\psi ={}& g \Big[ A^{+a}\bar{\psi}\gamma^- T^a \psi + A^{-a}\bar{\psi}\gamma^+ T^a \psi \nonumber \\
  &\phantom{aa}- \underbrace{A^{a}\bar{\psi}\bar{\gamma} T^a \psi}_{=0} - \bar{A}^{a}\bar{\psi}\gamma T^a \psi \Big]\,.
\end{align}
Note that we have also written the $\gamma$-matrices in the lightcone notation now, \textit{i.e.} $\gamma^+ \equiv \frac{1}{\sqrt{2}}(\gamma^0+\gamma^3)$,
\textit{etc.}
The expression for $\bar{A}^a$, following from eq.~\eqref{eom}, should then be replaced by
\begin{align}
  \bar{A}^a ={}& \frac{\partial^+}{\partial}A^{-\,a} + \frac{\partial^-}{\partial}A^{+\,a} \nonumber \\
  &- gf_{abc} \frac{1}{\partial^2} \left[ ((\partial A^{+\,b})A^{-\,c}) + ((\partial A^{-\,b})A^{+\,c}) \right] \nonumber \\
  &-g\frac{1}{\partial^2} \left( \bar{\psi} \gamma T^a \psi \right)\,,
\end{align}
and after substituting this back into the original Lagrangian we obtain
\begin{align}
  \mathcal{L} = \mathcal{L}_{YM} + \mathcal{L}_{q,0} + \mathcal{L}_{q,I} \,,
\end{align}
where
\begin{align}
  \mathcal{L}_{q,I} ={}& g\left[ A^{+a}\bar{\psi} \gamma^- T^a \psi -  \left(\frac{\partial^-}{\partial} A^{+a} \right) \bar{\psi}\gamma T^a \psi \right] \nonumber \\
  &+ g \left[ A^{-a}\bar{\psi} \gamma^+ T^a \psi -  \left(\frac{\partial^+}{\partial} A^{-a} \right) \bar{\psi}\gamma T^a \psi \right] \nonumber \\
  & + g^2 f_{abc} \left[ \frac{1}{\partial} \left( (\partial A^{+a}) A^{-c} \right) \frac{1}{\partial} \left( \bar{\psi}\gamma T^b\psi\right) \right. \nonumber \\
  &\phantom{AAAAAA}\left. - \frac{1}{\partial} \left( (\partial A^{-c}) A^{+a} \right) \frac{1}{\partial} \left( \bar{\psi}\gamma T^b\psi\right)\right] \nonumber \\
  & -\frac{g^2}{2} \frac{1}{\partial} \left( \bar{\psi}\gamma T^a \psi \right) \frac{1}{\partial}\left( \bar{\psi}\gamma T^a \psi \right)\,.
  \label{L_I}
\end{align}

\section{Feynman rules involving quarks\label{sec:QF}}
The external states for quarks are given by
\begin{eqnarray}
  \parbox{30mm}{
    \begin{fmffile}{ex.u.dirac}
      \begin{fmfgraph*}(60,40)
        \fmfleft{i1}
        \fmfright{o1}
        \fmf{fermion}{o1,i1}

        \fmfdot{o1}

        \fmflabel{$P$}{i1}
      \end{fmfgraph*}
    \end{fmffile}
  }
  \!\!\!\!\! =
  \bar{u}^s(P) \,,
\end{eqnarray}
and
\begin{eqnarray}
  \parbox{30mm}{
    \begin{fmffile}{ex.v.dirac}
      \begin{fmfgraph*}(60,40)
        \fmfleft{i1}
        \fmfright{o1}
        \fmf{fermion}{i1,o1}

        \fmfdot{o1}

        \fmflabel{$P$}{i1}
      \end{fmfgraph*}
    \end{fmffile}
  }
  \!\!\!\!\! =
  v^s(P) \,,
\end{eqnarray}
where $\bar{u}^s = (u^s)^{\dagger}\gamma_0$, and $u^s$ and $v^s$ are positive- and negativ-energy solutions, respectively, of the Dirac equation,
with $s$ labelling spin up or down.
The propagator is just the usual Dirac propagator
\begin{equation}
  \parbox{30mm}{
    \begin{fmffile}{dirac_prop}
      \begin{fmfgraph*}(60,40)
        \fmfleft{i1}
        \fmfright{o1}
        \fmf{fermion,label=$Q$}{o1,i1}

        \fmfdot{i1,o1}

      \end{fmfgraph*}
    \end{fmffile}
  }
  \!\!\!\!\! =
  \frac{\slashed{Q}+m}{Q^2-m^2}\,.
\end{equation}

From eq.~\eqref{L_I} we can read off the color-ordered Feynman rules involving vertices with quarks.

The three-point vertices involving one gluon and a quark-antiquark pair is
\\ \\
\begin{eqnarray}
  \parbox{30mm}{
    \begin{fmffile}{qqg}
      \begin{fmfgraph*}(70,50)
        \fmfbottom{i1,i2}
        \fmftop{o1}
        \fmf{fermion}{v1,i1}
        \fmf{fermion}{i2,v1}
        \fmf{gluon}{o1,v1}

        \fmflabel{$$}{i1}
        \fmflabel{$$}{i2}
        \fmflabel{$P^{\pm}$}{o1}
      \end{fmfgraph*}
    \end{fmffile}
  }
  = \phantom{a} \gamma^{\mp} - \frac{p^{\mp}}{p}\gamma \,,
\end{eqnarray}
\\
and
\\
\begin{eqnarray}
  \parbox{30mm}{
    \begin{fmffile}{qqg_rev}
      \begin{fmfgraph*}(70,50)
        \fmfbottom{i1,i2}
        \fmftop{o1}
        \fmf{fermion}{i1,v1}
        \fmf{fermion}{v1,i2}
        \fmf{gluon}{o1,v1}

        \fmflabel{$$}{i1}
        \fmflabel{$$}{i2}
        \fmflabel{$P^{\pm}$}{o1}
      \end{fmfgraph*}
    \end{fmffile}
  }
  = \phantom{a} -\left(  \gamma^{\mp} - \frac{p^{\mp}}{p}\gamma \right) \,.
  \label{qqg}
\end{eqnarray}
\\
or in case of a reference leg
\\ \\
\begin{eqnarray}
\parbox{30mm}{
\begin{fmffile}{qqg_ref}
     \begin{fmfgraph*}(70,50)
     \fmfbottom{i1,i2}
     \fmftop{o1}
     \fmf{fermion}{v1,i1}
     \fmf{fermion}{i2,v1}
     \fmf{gluon}{o1,v1}

     \fmflabel{$$}{i1}
     \fmflabel{$$}{i2}
     \fmflabel{$P^{\pm}_{ref}$}{o1}
     \end{fmfgraph*}
     \end{fmffile}
}
 = \phantom{a}  - \gamma \,,
\end{eqnarray}
\\
and
\\
\begin{eqnarray}
\parbox{30mm}{
\begin{fmffile}{qqg_rev_ref}
     \begin{fmfgraph*}(70,50)
     \fmfbottom{i1,i2}
     \fmftop{o1}
     \fmf{fermion}{i1,v1}
     \fmf{fermion}{v1,i2}
     \fmf{gluon}{o1,v1}

     \fmflabel{$$}{i1}
     \fmflabel{$$}{i2}
     \fmflabel{$P^{\pm}_{ref}$}{o1}
     \end{fmfgraph*}
     \end{fmffile}
}
 = \phantom{a} \gamma  \,.
\label{qqg_ref}
\end{eqnarray}

Due to the elimination of the $\bar{A}$ component from the Lagrangian, we have instead introduced four-point vertices involving quarks.
The first one is an effective gluon-gluon-quark-antiquark interaction
\\
\begin{eqnarray}
  \parbox{30mm}{
    \begin{fmffile}{qqgg}
      \begin{fmfgraph*}(70,50)
        \fmfleft{i1,i2}
        \fmfright{o1,o2}
        \fmf{fermion}{v1,i1}
        \fmf{gluon}{i2,v1}
        \fmf{fermion}{o1,v1}
        \fmf{gluon}{o2,v1}

        \fmflabel{$Q$}{i1}
        \fmflabel{$K^{\pm}$}{i2}
        \fmflabel{$P^{\mp}$}{o2}
        \fmflabel{$T$}{o1}
      \end{fmfgraph*}
    \end{fmffile}
  }
  = \phantom{a}  \frac{(p-k)}{(k+p)^2}\gamma \,,
  \label{qqgg}
\end{eqnarray}
\\ 
and
\\ 
\begin{eqnarray}
  \parbox{30mm}{
    \begin{fmffile}{qqgg2}
      \begin{fmfgraph*}(70,50)
        \fmfleft{i1,i2}
        \fmfright{o1,o2}
        \fmf{fermion}{i1,v1}
        \fmf{gluon}{i2,v1}
        \fmf{fermion}{v1,o1}
        \fmf{gluon}{o2,v1}

        \fmflabel{$Q$}{i1}
        \fmflabel{$K^{\pm}$}{i2}
        \fmflabel{$P^{\mp}$}{o2}
        \fmflabel{$T$}{o1}
      \end{fmfgraph*}
    \end{fmffile}
  }
  = \phantom{a} - \frac{(p-k)}{(k+p)^2}\gamma \,.
  \label{qqgg2}
\end{eqnarray}
Like in the pure gluon case there can not be reference legs on these four-point vertices either. The second one is a double pair quark-antiquark interaction
\\
\begin{equation}
  \parbox{30mm}{
    \begin{fmffile}{qqqq1}
      \begin{fmfgraph*}(70,50)
        \fmfleft{i1,i2}
        \fmfright{o1,o2}
        \fmf{fermion}{v1,i1}
        \fmf{fermion}{i2,v1}
        \fmf{fermion}{o1,v1}
        \fmf{fermion}{v1,o2}

        \fmflabel{$Q$}{i1}
        \fmflabel{$K$}{i2}
        \fmflabel{$P$}{o2}
        \fmflabel{$T$}{o1}
      \end{fmfgraph*}
    \end{fmffile}
  }
  = \phantom{a} \frac{\gamma}{(t+q)}\frac{\gamma}{(k+p)} + \frac{\gamma}{(q+k)}\frac{\gamma}{(p+t)} \,,
  \label{qqqq1}
\end{equation}
\\
and
\\
\begin{equation}
  \parbox{30mm}{
    \begin{fmffile}{qqqq2}
      \begin{fmfgraph*}(70,50)
        \fmfleft{i1,i2}
        \fmfright{o1,o2}
        \fmf{fermion}{i1,v1}
        \fmf{fermion}{v1,i2}
        \fmf{fermion}{o1,v1}
        \fmf{fermion}{v1,o2}

        \fmflabel{$Q$}{i1}
        \fmflabel{$K$}{i2}
        \fmflabel{$P$}{o2}
        \fmflabel{$T$}{o1}
      \end{fmfgraph*}
    \end{fmffile}
  }
  = \phantom{a} -\frac{\gamma}{(q+k)}\frac{\gamma}{(p+t)} \,.
  \label{qqqq2}
\end{equation}
\\
Notice that the $\gamma$-matrices in eq.~\eqref{qqqq1} and \eqref{qqqq2} are \textit{not} multiplied together. In a
real amplitude calculation they will appear between corresponding external spinors, for example like
\\
\begin{eqnarray}
  \parbox{30mm}{
    \begin{fmffile}{qqqq1_sp}
      \begin{fmfgraph*}(70,50)
        \fmfleft{i1,i2}
        \fmfright{o1,o2}
        \fmf{fermion}{i1,v1}
        \fmf{fermion}{v1,i2}
        \fmf{fermion}{v1,o1}
        \fmf{fermion}{o2,v1}

        \fmflabel{$1$}{i1}
        \fmflabel{$2$}{i2}
        \fmflabel{$3$}{o2}
        \fmflabel{$4$}{o1}
      \end{fmfgraph*}
    \end{fmffile}
  } &=& \frac{\bar{u}(P_2)\gamma v(P_1)}{p_1 + p_2} \frac{\bar{u}(P_4)\gamma v(P_3)}{p_4 + p_3} \nonumber \\
  &&+ (2 \leftrightarrow 4)\,.
\end{eqnarray}

This covers all the Feynman rules one obtains for QCD in the space-cone gauge.

\section{Conclusions\label{sec:con}}
In this paper we have explicitly written down all Feynman rules for QCD in space-cone gauge when unphysical degrees of freedom in the gluonic sector
have been removed. Combined with a clever choice of reference frame this reduces the amount of Feynman diagrams needed
for gluon amplitude calculations considerably. We then made some comments about the close connection between BCFW recursion relations
and the space-cone gauge, especially concerning the role played by the four-point vertex. We have also seen that in the presence of quarks the former manipulations of the Lagrangian lead to the introduction of effective four-point interaction terms involving quark-antiquark pairs. 
\begin{acknowledgments}
  \noindent We would like to thank Poul Henrik Damgaard, Emil Bjerrum-Bohr and Diana Vaman for useful discussions and comments. TS would also like to thank Simon Badger for helpful comments.
\end{acknowledgments}

\end{document}